\documentclass[12pt,preprint]{aastex}

\catcode`\@=11
\newcommand{\gapprox}{\mathrel{\mathpalette\@versim>}}
\newcommand{\lapprox}{\mathrel{\mathpalette\@versim<}}
\newcommand{\propapprox}{\mathrel{\mathpalette\@versim\propto}}
\newcommand{\@versim}[2]
  {\lower3.1truept\vbox{\baselineskip0pt\lineskip0.5truept
\ialign{$\m@th#1\hfil##\hfil$\crcr#2\crcr\sim\crcr}}}
\catcode`\@=12
\shorttitle{DUST DESTRUCTION IN LMC TYPE Ia SNRS}
\shortauthors{BORKOWSKI ET AL.}

\begin{document}

\title{Dust Destruction in Type Ia Supernova Remnants in 
the Large Magellanic Cloud}

\author{Kazimierz J. Borkowski,\altaffilmark{1}
Brian J. Williams,\altaffilmark{1}
Stephen P. Reynolds,\altaffilmark{1}
William P. Blair,\altaffilmark{2}
Parviz Ghavamian,\altaffilmark{2}
Ravi Sankrit,\altaffilmark{2}
Sean P. Hendrick,\altaffilmark{3}
Knox S. Long,\altaffilmark{4}
John C. Raymond,\altaffilmark{5}
R. Chris Smith,\altaffilmark{6}
Sean Points,\altaffilmark{6}
\& P. Frank Winkler\altaffilmark{7}
}

\altaffiltext{1}{Physics Dept., North Carolina State U.,
    Raleigh, NC 27695-8202; kborkow@ncsu.edu, steve\_reynolds@ncsu.edu,
    bjwilli2@ncsu.edu}
\altaffiltext{2}{Dept. of Physics and Astronomy, Johns Hopkins, 
    3400 N. Charles St., Baltimore, MD 21218-2686; wpb@pha.jhu.edu, 
parviz@pha.jhu.edu, ravi@pha.jhu.edu}
\altaffiltext{3}{Physics Dept., Millersville U., PO Box 1002, Millersville, 
    PA 17551; sean.hendrick@millersville.edu}
\altaffiltext{4}{STScI, 3700 San Martin Dr., Baltimore, MD 21218; 
    long@stsci.edu}
\altaffiltext{5}{Harvard-Smithsonian Center for Astrophysics, 60 Garden
    Street, Cambridge, MA 02138; jraymond@cfa.harvard.edu}
\altaffiltext{6}{CTIO, Cailla 603, La Serena, Chile; csmith@noao.edu, 
    spoints@noao.edu}
\altaffiltext{7}{Dept. of Physics, Middlebury College, Middlebury, VT 
    05753; winkler@middlebury.edu}

\begin{abstract}

We present first results from an extensive survey of Magellanic Clouds
supernova remnants (SNRs) with the {\sl Spitzer Space Telescope}.  We
describe IRAC and MIPS imaging observations at 3.6, 4.5, 5.8, 8, 24,
and 70 $\mu$m of four Balmer-dominated Type Ia SNRs in the Large
Magellanic Cloud (LMC): DEM L71 (0505-67.9), 0509--67.5, 0519--69.0,
and 0548-70.4.  None was detected in the four short-wavelength IRAC
bands, but all four were clearly imaged at 24 $\mu$m, and two at 70
$\mu$m.  A comparison of these images to {\sl Chandra} broadband X-ray
images shows a clear association with the blast wave, and not with
internal X-ray emission associated with ejecta.  Our observations are
well described by 1-D shock models of collisionally heated dust
emission, including grain size distributions appropriate for the LMC,
grain heating by collisions with both ions and electrons, and
sputtering of small grains.  Model parameters are constrained by
X-ray, optical, and far-ultraviolet observations.  Our models can
reproduce observed 70/24 $\mu$m flux ratios only by including
sputtering, destroying most grains smaller than 0.03--0.04 $\mu$m in 
radius. We infer total
dust masses swept up by the SNR blast waves, before sputtering, of
order $10^{-2} \ M_\odot$, several times less than those
implied by a dust/gas mass ratio of 0.3\% as often assumed for the
LMC.  Substantial dust destruction has implications for gas-phase
abundances.

\end{abstract}


\section{Introduction}
\label{intro}

The dust content in galaxies, dust composition, and grain size
distribution are determined by the balance between dust formation,
modification in the interstellar medium (ISM), and destruction
\citep{draine03}.  Some evidence exists for dust formation in the
ejecta of core-collapse supernovae \citep[e.g., SN 1987A;][]{dekool98}
but no reports exist for SNe Ia.  Dust {\it destruction} is
intrinsically linked to SN activity, through sputtering in gas heated
by energetic blast waves and through betatron acceleration in
radiative shocks \citep{jones04}.  Dust destruction in SNRs can be
studied by its strong influence on thermal IR emission from
collisionally heated dust.  The IRAS All Sky Survey provided
fundamental data on Galactic SNRs \citep{arendt89,saken92}. This
prompted extensive theoretical work on dust heating, emission, and
destruction within hot plasmas, summarized by
\citet{dwekarendt92}. Theory is broadly consistent with IRAS
observations, but limitations of those observations (low spatial and
spectral resolution and confusion with the Galactic IR background)
precluded any detailed comparisons. In particular, while it is clear
that thermal dust emission is prevalent in SNRs, our understanding of
dust destruction is quite poor.

To examine the nature of dust heating and destruction in the
interstellar medium, we conducted an imaging survey with the {\sl
Spitzer Space Telescope} (SST) of 39 SNRs in the Magellanic Clouds
(MCs).  We have selected a subset of our detections, four remnants of
Type Ia supernovae, to address questions of dust formation in Type Ia
ejecta, dust content of the diffuse ISM of the LMC, and dust
destruction in SNR shocks.  Both DEM L71
(0505-67.9; Rakowski, Ghavamian, \& Hughes 2003) and 
0548--70.4 (Hendrick, Borkowski, \& Reynolds 2003) show
X-ray evidence for iron-rich ejecta in the interior, and both have
well-studied Balmer emission from nonradiative shocks
\citep{ghavamian03,smith91}.  Two smaller remnants, 0509--67.5 and
0519--69.0, also show prominent H$\alpha$ and Ly$\beta$ emission from
nonradiative shocks (Tuohy et al.~1982; Smith et al.~1991; Ghavamian
et al. 2006, in preparation).  There appears to be little or no
optical contribution from radiative shocks.  Confusion in IR is
widespread in the LMC, but our remnants are less confused than
typical, easing the task of separating SNR emission from background.

\section{Observations and Data Reduction}
\label{obs}

\begin{figure*}
\centering
\includegraphics[width=14.5cm, angle=0]{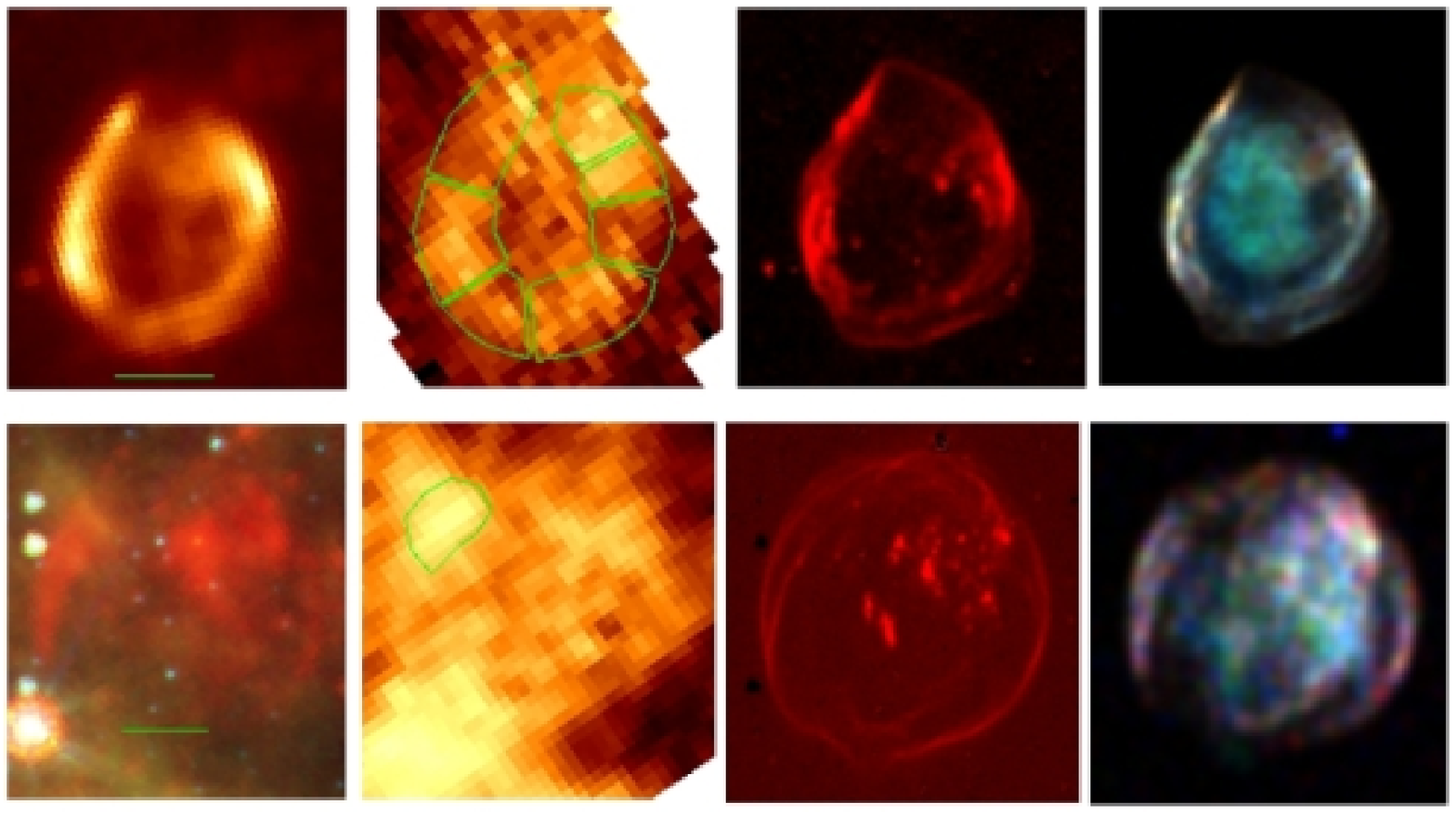}
\includegraphics[width=14.5cm, angle=0]{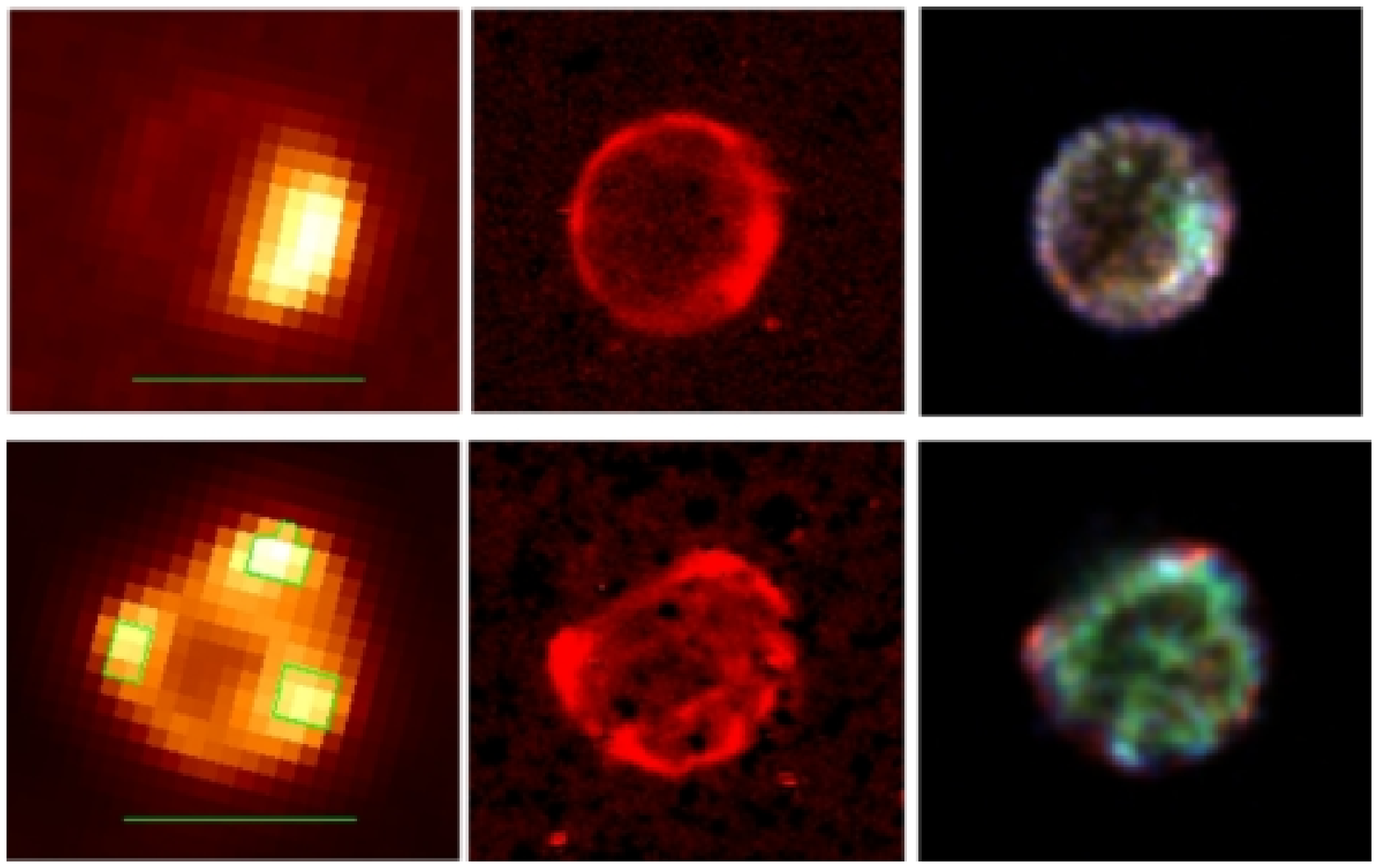}
\caption{Top row:  DEM L71 at 24 and 70 $\mu$m,
H$\alpha$, and X-ray (red, 0.3 -- 0.7 keV; green,
0.7 -- 1.0 keV; blue, 1.0 -- 3.5 keV; smoothed with
1 pixel Gaussian).
Second row: 0548--70.4 with red, 24 $\mu$m; green, IRAC 8.0 $\mu$m; 
blue $\mu$m, IRAC 5.8 $\mu$m; 70 $\mu$m; H$\alpha$, and X-ray image
as for DEM L71, smoothed with a 2 pixel Gaussian. 
Third row:  0509--67.5 at 24 $\mu$, 
H $\alpha$, and X-ray: red, 0.3 -- 0.7 keV; green, 0.7 -- 1.1 keV and blue, 
1.1 -- 7.0 keV, 
Fourth row:  0519--69.0, as in third row. Half-arcmin scales are shown for each SNR.
\label{images}
}
\end{figure*}

We observed all four objects in all four bands of the Infrared Array
Camera (IRAC), as well as with the Multiband Imaging Photometer for
Spitzer (MIPS) at 24, 70, and 160 $\mu$m. The 160 $\mu$m images contain 
only emission from the general ISM, so we do not discuss them here. 
Each IRAC observation totaled 300
s (10 30-s frames); at 24 $\mu$m, 433 s total (14 frames); and at 70
$\mu$m, 986 s total (94 frames) for all but 0548--70.4, for which we
observed a total of 546 s in 52 frames.  The observations took place
between November 2004 and April 2005.  Images are shown in
Figure~\ref{images}.  Confusion from widely distributed warm
dust made many 70 $\mu$m observations problematic, but we obtained
useful data on both DEM L71 and 0548--70.4. 

MIPS images were processed from Basic Calibrated Data (BCD) to
Post-BCD (PBCD) by v.~11 of the SSC PBCD pipeline. For the 24 $\mu$m
images, we then re-mosaicked the stack of BCD images into a PBCD
mosaic using the SSC-provided software MOPEX, specifically the overlap
correction, to rid the images of artifacts.  For the 70 $\mu$m data,
we used the contributed software package GeRT, provided also by the
SSC, to remove some vertical streaking.  IRAC images were also
reprocessed using MOPEX to rid the image of artifacts caused by bright
stars.

All four remnants were clearly detected at 24 $\mu$m, with fluxes
from indicated regions
reported in Table~\ref{fluxtable}.  As Figure~1 shows, emission is
clearly associated with the X-ray-delineated blast wave, though not
with interior X-ray emission.  Since we expect line emission from
fine-structure transitions of low-ionization material to be a
significant contributor only in cooler, denser regions identified by
radiative shocks, we conclude that the emission we detect is
predominantly from heated dust.  None of our objects was clearly
detected at 8 $\mu$m, with fairly stringent upper limits shown in
Table~\ref{fluxtable}.

\section{Discussion}
\label{disc}

We modeled the observed emission assuming collisionally heated dust
\citep[e.g.,][]{dwekarendt92}.  The models allow an arbitrary
grain-size distribution, and require as input parameters the hot gas
density $n$, electron temperature $T_e$, ion temperature $T_i$, and
shock sputtering age $\tau=\int_0^t n_p dt$.  The models use an
improved version of the code described by \citet{borkowski94},
including a method devised by \citet{guh89} to account for
transiently-heated grains, whose temperature fluctuates with time and
therefore radiate far more efficiently.  The energy deposition rates
by electrons and protons were calculated according to \citet{dwek87} and
\citet{dweksmith96}.  We used dust emissivities based on bulk optical
constants of \citet{drainelee84}.  Our non-detections in IRAC bands
showed that small grains are destroyed, so it was not necessary to
model emission features from small polycyclic aromatic hydrocarbon
(PAH) grains.  The preshock grain size distribution was taken from the
``provisional'' dust model of \citet{wg01}, consisting of separate
carbonaceous and silicate grain populations, in particular their
average LMC model with maximal amount of small carbonaceous grains.
Sputtering rates are based on sputtering cross sections of
\citet{bianfer05}, augmented by calculations of an enhancement in
sputtering yields for small grains by \citet{jurac98}. We have modeled
1-D shocks, that is, superposed emission from regions of varying
sputtering age from zero up to a specified shock age \citep{dwek96}.

To estimate shock parameters, we used the non-radiative shock models
of Ghavamian et al.~(2001) to model the broad component H$\alpha$
widths and broad-to-narrow H$\alpha$ flux ratios measured by
\cite{tuohy82} and \cite{smith91} for the LMC SNRs.  Results for
electron and proton temperatures $T_e$ and $T_p$ are quoted in
Table~\ref{inputs}.  For 0509--67.5, we assumed $T_e/T_p \le 0.1$ at
the shock front, consistent with the observed Ly$\beta$ FWHM of 3700
km s$^{-1}$ (Ghavamian et al. 2006, in preparation).  For Sedov
dynamics, the sputtering age $\tau$ (which is also the ionization
timescale) reaches a maximum of about $(1/3)n_p t$ where $t$ is the
true age of the blast wave \citep{borkowski01}.  Therefore we use an
``effective sputtering age'' of $n_p t/3$ when calculating effects of
sputtering.

\subsection{DEM L71 and 0548--70.4}

These two remnants have been well-studied in X-rays (DEM L71: Rakowski
et al.~2003; 0548--70.4, Hendrick et al.~2003).  They have ages of
4400 and 7100 yr, respectively, derived from Sedov models.  For DEM
L71, Ghavamian et al.~(2003) were able to infer
shock velocities over much of the periphery, ranging from
430 to 960 km s$^{-1}$, 
consistent with X-ray inferences \citep{rakowski03}.

To model DEM L71, we used parameters deduced from {\sl Chandra}
observations \citep{rakowski03}, averaged over the entire blast wave
since different subregions were fairly similar.  We find a predicted
70/24 ratio in the absence of sputtering ($\tau = 0$) of about 2.3 
(including only grains larger than 0.001 $\mu$m in radius),
compared to the observed 5.1.  Using an effective age of 1/3 the Sedov
age gave a value of 5.1.  Table~\ref{results} also gives the total
dust mass we derive, and the total IR luminosity produced by the
model.  

For 0548--70.4, both the east and west limbs and some bright knots of
interior emission are visible at 24 $\mu$m, but only the north half of
the east limb is clearly detected at 70 $\mu$m. Only fluxes from this
region were measured; the results are summarized in
Table~\ref{fluxtable}.  Using a 1-D model for Coulomb heating of
electrons by protons, we calculate a mean electron temperature in the
shock region of $T_e \sim 0.66$ keV.  A model using the postshock
density of 0.72 cm$^{-3}$ obtained by Hendrick et al.~(2003) for the
whole limb (including sputtering) gives too high a 70/24 $\mu$m
ratio. That ratio is very sensitive to density; we found that
increasing $n_p$ by a factor $< 2.5$ adequately reproduced the observed
ratio.  That fitted density appears in Table~\ref{inputs} and the
corresponding results are in Table~\ref{results}. Gas mass was derived
from the X-ray emission measure of the east limb \citep{hendrick03}, 
scaled to the region shown in Figure~1, and using electron density in 
Table~\ref{inputs}.

\subsection{0509--67.5 and 0519--69.0}

Our other two objects are much smaller; X-ray data suggest young ages
\citep{warren04}.  Detections of light echoes \citep{rest05} indicate
an age of about 400 yr for 0509--67.5 and about 600 yr for 0519--69.0,
with $\sim 30$\% errors.  Much higher shock velocities inferred by
Ghavamian et al.~(2006, in preparation) mean that plasma heating
should be much more effective.  Higher dust temperatures, hence lower
70/24 $\mu$m ratios, should result.  In fact, we did not detect either
remnant at 70 $\mu$m, with upper limits on the ratio considerably
lower than the other two detections (Table~\ref{fluxtable}).

In the case of 0509-67.5, optical-UV observations fix only $T_p$, so
we regarded the density $n_p$ as a free parameter, fixing $\tau$ at
$n_p t/3$ and finding $T_e$ assuming no collisionless heating.  Our 70
$\mu$m upper limit gives a lower limit on $n_p$, shown in
Table~\ref{inputs}, as well as an upper limit on the total dust mass
(Table~\ref{results}).

The analysis of 0519-69.0 was identical to that done for 0509-67.5. 
However, for 0519-69.0 we divided the remnant up
into two regions: the three bright knots (which we added together and
considered one region, accounting for 20\% of the total flux) and the
rest of the blast wave.  Optical spectroscopy (Ghavamian et al.~2006)
allowed determination of parameters separately for the knots and the
remainder.  Again regarding $n_p$, $T_e$ and $\tau$ as free parameters,
we place lower limits on the post-shock densities and $T_e$, and upper
limits on the amount of dust mass, including sputtering.  For both
remnants, density limits assume the effective sputtering age; if there
is no sputtering at all, we obtain firm lower limits on density lower
by less than a factor of 2.

\section{Results and Conclusions}
\label{concls}

The IR emission in the Balmer-dominated SNRs in the LMC is spatially
coincident with the blast wave. It is produced within the shocked ISM
by the swept-up LMC dust heated in collisions with thermal electrons
and protons.  We find no evidence for infrared emission associated
with either shocked or unshocked ejecta of these thermonuclear
SNRs. While detailed modeling of small grains is required to make a
quantitative statement, apparently little or no dust forms in such
explosions, and any line emission produced by ejecta is below our
detection limit. This is consistent with observations of Type Ia SNe
where dust formation has never been observed.  It is also consistent
with the absence in meteorites of presolar grains formed in Type Ia
explosions \citep{clayton04}.

The measured 70/24 $\micron$ MIPS ratios in DEM L71 and 0548-70.4, and
the absence of detectable emission in the IRAC bands in all 4 SNRs,
can be accounted for with dust models which include destruction of
small grains. Without dust destruction, numerous small grains present
in the LMC ISM \citep[e.g.,][]{wg01} would produce too much emission
at short wavelengths when transiently heated to high temperatures by
energetic particles. Destruction of small grains is required to reproduce 
the observed 70/24 $\micron$ MIPS ratios in DEM L71 and 0548-70.4: 
90\%\ of the mass in grains smaller than 0.03--0.04 $\mu$m is destroyed in
our models.
Even with this destruction, we
infer pre-sputtering dust/gas mass far smaller than the 0.25\% in the
Weingartner \& Draine model.

The two young remnants, 0509--67.5 and 0519--69.0, have been detected
only at 24 $\micron$, but our rather stringent upper limits at 70
$\micron$ suggest the presence of much hotter dust than in the older
SNRs DEM L71 and 0548-70.4. Such hot dust is produced in our plane
shock models only if the postshock electron densities exceed $1.6$
cm$^{-3}$ and 3.4--7.7 cm$^{-3}$ in 0509--67.5 and 0519--69.0,
respectively (Table~\ref{inputs}).  0509--67.5 is asymmetric, and the
quoted lower density limit needs to be reduced if an average postshock
electron density representative of the whole SNR is of interest. We
measure a flux ratio of 5 between the bright and faint hemispheres,
depending primarily on the gas density ratio between the hemispheres,
and on the ratio of swept-up ISM masses. For equal swept-up masses,
our models reproduce the observed ratio for a density contrast of 3 or
less; the actual density contrast is lower because more mass has been
swept up in the brighter hemisphere. The nearly circular shape of
0509--67.5 also favors a low density contrast.  The densities derived
here are several times higher than an upper limit to the postshock
density of $0.2$ cm$^{-3}$ obtained by \citet{warren04} who used
hydrodynamical models of \citet{dwarkadas98} to interpret {\it
Chandra} X-ray observations of this SNR. The origin of this
discrepancy is currently unknown.  Possible causes include: (1)
neglect of extreme temperature grain fluctuations in our dust models
for 0509--67.5, (2) modification of the blast wave by cosmic rays as
suggested for the Tycho SNR by \citet{warren05}, (3) contribution of
line emission in the 24 $\micron$ MIPS band. 
 
The measured 24 and 70 $\micron$ IR fluxes, in combination with
estimates of the swept-up gas from X-ray observations, imply a
dust/gas ratio a factor of several lower than typically assumed for
the LMC. In order to resolve this discrepancy, one needs much higher
dust destruction rates and/or a much lower dust/gas ratio in the
pre-shock gas. Most determinations of dust mass come from
higher-density regions, but Type Ia SNRs are generally located in the
diffuse ISM, where densities are low. Both the dust content and the
grain size distribution might be different in the diffuse ISM. In the
Milky Way, the dust content is lower in the more diffuse ISM
\citep[e.g.,][]{savage96}, most likely due to dust destruction by
sputtering in fast SNR shocks (more prevalent at low ISM densities)
and by grain-grain collisions in slower radiative shocks.
Grain-grain collisions are the more likely destruction mechanism for
large grains \citep{jones94,borkowski95}, so such grains might be 
less common in the diffuse ISM.  Smaller grains are more efficiently
destroyed by sputtering in SNRs, so dust destruction will be more
efficient for a steeper preshock grain size distribution (more
weighted toward small grains).  This in combination with the lower
than average preshock dust content mostly likely accounts for the
observed deficit of dust in the Balmer-dominated SNRs in the
LMC. Apparently dust in the ambient medium near these SNRs has been
already affected (and partially destroyed) by shock waves prior to its
present encounter with fast SNR blast waves.  Spectroscopic follow-up
is required in order to confirm preliminary conclusions presented in this work 
and learn more about dust and its destruction in the diffuse ISM of the LMC.

\acknowledgments

We thank Joseph Weingartner and Karl Gordon for discussions about dust
in the LMC.
This work was supported by NASA through Spitzer Guest Observer grant
RSA 1265236, and by {\sl Chandra} Archival Research grant SAO
AR5-6007X.  PFW acknowledges NSF grant AST-0307613.



\begin{deluxetable}{lccc}
\tablecolumns{4}
\tablewidth{0pc}
\tabletypesize{\footnotesize}
\tablecaption{Measured Fluxes and Upper Limits\tablenotemark{a}}
\tablehead{
\colhead{Object} & 8.0 $\mu$m & 24 $\mu$m & 70 $\mu$m}

\startdata
DEM L71 & $<1.06$ & 88.2 $\pm${8.8} & 455 $\pm${94}\\
0548-70.4 & $<3.82$ & 2.63 $\pm${0.30} & 19.9 $\pm${4.7}\\
0509-67.5 & $<0.2$ & 16.7 $\pm${1.7} & $<32.7$\\
0519-69.0 & $<0.9$ & 92.0 $\pm${9.2} & $<121$\\

\enddata

\tablenotetext{a}{All fluxes (not color-corrected) in mJy.  Limits are $3\sigma$.}
\label{fluxtable}
\end{deluxetable}


\begin{deluxetable}{lccccccc}

\tablecolumns{4}
\tablewidth{0pc}
\tabletypesize{\footnotesize}
\tablecaption{Model Input Parameters}
\tablehead{
\colhead{Object} & $T_e$ (keV) & $T_p$ (keV) & $n_p$ & $n_e$
& Age (yrs.)& $\tau (10^{10}$ cm$^{-3}$ s) & Ref.}

\startdata
DEM L71                    & 0.65    & 1.1 & 2.3    & 2.7    & 4400 & 11   & 1, 2\\
0548-70.4                  & 0.65    & 1.5 & 1.7    & 2.0    & 7100 & 12   & 3, 4\\
0509-67.5                  & 1.9     & 89  & $>1.4$ & $>1.6$ & 400  & 0.59 & 4, 5\\
0519-69.0\tablenotemark{a} & 2.1     & 36  & $>2.8$ & $>3.4$ & 600  & 1.8  & 4, 5\\
0519-69.0\tablenotemark{b} & 1.0     & 4.2 & $>6.4$ & $>7.7$ & 600  & 4.0  & 4, 5\\

\enddata

\tablenotetext{a}{Fainter portions of remnant}
\tablenotetext{b}{Three bright knots}

\tablecomments{Densities are post-shock.  References: (1) Rakowski et al 2003; 
(2) Ghavamian et al 2003; (3) Hendrick et al 2003; 
(4) Ghavamian et al 2006, in preparation; (5) Rest et al. 2005}
\label{inputs}
\end{deluxetable}

\begin{deluxetable}{lcccccccc}
\tablecolumns{4}
\tablewidth{0pc}
\tabletypesize{\footnotesize}
\tablecaption{Model Results}
\tablehead{
\colhead{Object} & 70/24 (0) & 70/24 sput. & 70/24 obs. & $T$(dust)(K) 
& Dust Mass 
& \% destr. 
& dust/gas 
& $L_{36}$ }

\startdata
DEM L71                    & 2.3       & 5.1         & 5.1        & 55--65   
   & 0.034                   & 35                & 4.2$\times 10^{-4}$    & 12\\
0548-70.4                  & 2.7       & 7.6         & 7.6        & 53--62   
   & 0.0018                & 40                & 7.5$\times 10^{-4}$  & 2.1\\
0509-67.5                  & $<2.0$    & $<2.0$      & $<2.0$     & 66--70   
   & $<1.1 \times 10^{-3}$    & $>18$                & ... & ...\\
0519-69.0\tablenotemark{a} & $<1.3$    & $<1.3 $     & $<1.3$     & 72--77  
   & $<2.7 \times 10^{-3}$    & $>34$                & ... & ...\\
0519-69.0\tablenotemark{b} & $<1.3$    & $<1.3 $     & $<1.3$     & 73--86  
   & $<6.4 \times 10^{-4}$    & $>38$                & ...                   & ...\\

\enddata

\tablenotetext{a}{Fainter portions of remnant}
\tablenotetext{b}{Three bright knots}
\tablecomments{Column 2: model prediction without sputtering; column 3, including
sputtering with $\tau = n_p t/3$; column 4, observations; column 5, for 0.02--0.1 $\mu$m grains; column 6, mass of dust
currently observed (after sputtering), in $M_\odot$; column 7, percentage of original
dust destroyed; column 8, ratio of swept-up dust to gas masses; column 9,
$L_{36} \equiv L_{IR}/10^{36}$ erg s$^{-1}$.}

\label{results}
\end{deluxetable}


\end{document}